\documentclass[twocolumn,showpacs,preprintnumbers,amsmath,amssymb]{revtex4}

\usepackage{graphicx}
\usepackage{dcolumn}
\usepackage{bm}

\begin{document}

\title{Thermal noise limitations to force measurements with torsion pendulums:
Applications to the measurement of the Casimir force and its
thermal correction LA-UR-04-4906}

\author{S.K. Lamoreaux and W.T. Buttler}

\affiliation{University of California,Los Alamos National
Laboratory,Physics Division P-23, M.S. H803, Los Alamos, NM 87545}

\date{July 14, 2004}

\begin{abstract}
A general analysis of thermal noise in torsion pendulums is
presented.  The specific case where the torsion angle is kept
fixed by electronic feedback is analyzed.  This analysis is
applied to a recent experiment that employed a torsion pendulum to
measure the Casimir force.  The ultimate limit to the distance at
which the Casimir force can be measured to high accuracy is
discussed, and in particular the prospects for measuring the
thermal correction are elaborated upon.

\end{abstract}
 \maketitle

\section{Introduction}

In a recent measurement of the Casimir force \cite{cas}, a torsion
pendulum was used.  This measurement provided the highest accuracy
measurements to date of this force at separations greater than one
$\mu$m \cite{lam}. This is because the torsion pendulum is among
the most sensitive physical measurment devices, as has been
recognized since Cavendish's measurement of the gravitational
constant $G$ over two centuries ago \cite{caven}.

The limitations to force measurements due to thermal energy in the
torsion mode, although long-known \cite{ising} and extensively
studied, the particular case as in \cite{lam} where electronic
feedback was employed to keep the angle fixed has, to our
knowledge, not been addressed at all. The interest in Casimir
force measurements suggests that we present the results of our
analysis of the experiment described in \cite{lam}.

Further measurements of the Casimir force, at large plate
separation, are motivated by recent theoretical work that suggests
a large correction to the thermal contribution to the Casimir
force \cite{bos}.  This work is not universally accepted
\cite{mostep} and it has been suggested that the correction is
material dependent \cite{lam2}. The specific point of \cite{lam2}
is that both \cite{bos} and \cite{mostep} are correct respectively
for dielectrics and for metals.  The theory put forward in
\cite{lam2} can be tested with a poorly conducting material such
as lightly doped germanium where the electron mean free path is
less than the electromagnetic skin depth in the material.  In this
case, the analysis presented in \cite{bos} is applicable.  For a
metal film as used in \cite{lam}, the analysis presented in
\cite{mostep} is applicable.  A comparison of the Casimir force
for metals, semiconductors, and dielectrics at distances greater
than 4 $\mu$m, with 10\% level accuracy, is necessary to test the
assertion put forward in \cite{lam2}.

In this note, we review the calculations of thermal noise in force
measurements for free torsion pendulums. Previously, R. Newman
\cite{newman} discussed three modes of operation of a torsion
pendulum;  here we analyze a fourth mode where magnetic or
electrostatic feedback is used to measure the force. Magnetic
feedback was employed in a demonstration experiment to measure the
gravitational constant \cite{crandall}, while this technique was
slightly modified in \cite{lam} where electrostatic feedback was
employed to eliminate non-linear magnetic and hysteresis effects.

\section{Angle Noise in a Free Torsion Pendulum}

We consider the case of a torsion pendulum where a mass $m$ with
moment of inertia $I$ is suspended in the Earth's gravitational
field by a fiber (or wire) with angular restoring torque
$\tau=-\alpha\theta$ at an absolute temperature $T$. Generally,
the thermal angular fluctuations for the swinging (gravitational)
pendulum modes are vastly smaller than the torsional angular
fluctuations.  Because each mode has $kT/2$ of thermal energy, the
RMS angular fluctuations are \cite{ising}
\begin{equation}
\delta\theta=\sqrt{kT\over \alpha}<< \sqrt{kT\over mg \ell}
\end{equation}
where $k$ is Boltzmann's constant, $g$ is the gravitational
acceleration, and $\ell$ is the pendulum length. For $\alpha= 1$
dyne/rad, $m=100$ g, and $\ell=10$ cm, the swinging mode thermal
noise amplitude is about three orders of magnitude smaller than
the torsion mode noise.

The case of the ``free" torsion pendulum is when there is no
external driving or restoring torque applied to the system. The
angular fluctuations, through the fluctuation-dissipation theorem
\cite{ll}  and the Langevin equation \cite{reif}, are describe by
\begin{eqnarray}
I\ddot \theta+\gamma \dot\theta +\alpha\theta
&=&\tau(t)\Rightarrow
\\
(-I\omega^2+i\gamma\omega+\alpha)\theta_\omega&=& \sqrt{4kT\gamma
\Delta f}
\end{eqnarray}
where $\tau(t)$ is the thermal fluctuation torque on the torsion
pendulum, and $\theta_\omega$ is the fluctuation amplitude at
frequency $\omega$ in a bandwidth $\Delta f$ Hz.

The mean square spectral density of the fluctuations is therefore
\begin{equation}
{\vert\theta_\omega\vert^2\over \Delta
f}={4kT\gamma\over\alpha^2}{1\over (1-x^2)^2+Q^{-2}x^2}
\end{equation}
where
\begin{equation}
x={\omega\over \omega_0};\ \ \ \omega_0^2={\alpha\over I}
\end{equation}
and the mechanical quality factor is
\begin{equation}
Q={\alpha\over \omega_0 \gamma}.
\end{equation}

To determine the variance of a series of measurements of $\theta$
we need the autocorrelation function $R_\theta(t)$.  This can be
determined by the inverse of the Wiener-Khinchin theorem
\cite{vdz},
\begin{equation}
R_\theta(t)=\Re\bigg[ \int _0^\infty {\vert
\theta_\omega\vert^2\over \Delta f}e^{i\omega t} d f\bigg]
\end{equation}
where $\Re$ means the real part and $\omega=2\pi f$.  This
integral can be calculated by contour integration methods by
noting that the denominator can be factored as
\begin{eqnarray}\label{denom}
(1-x^2)^2+Q^{-2}x^2&=&\nonumber\\
(x-\Omega+iQ^{-1}/2)&\times &(x+\Omega+iQ^{-1}/2)\times\nonumber\\
(x-\Omega-iQ^{-1}/2)&\times&(x+\Omega-iQ^{-1}/2)
\end{eqnarray}
where
\begin{equation}
\Omega=\sqrt{1-{Q^{-2}\over 4}}.
\end{equation}

Noting that $\vert\theta_\omega\vert^2=
\vert\theta_{-\omega}\vert^2$ and $df=\omega_0 dx/2\pi$, the
integration path is taken from $x=-\infty$ to $x=+\infty$, and the
contour is closed by a semicircle in the upper half-plane where
the poles are located at $\pm\Omega + iQ^{-1}/2$ so that
$\lim_{t\rightarrow \infty}R_\theta(t)=0$. We thus find
\begin{eqnarray}
R_\theta(t)&=&\Re \bigg[{4kT\gamma\over \alpha^2}{\omega_0\over
2\pi}{\pi\over 4}e^{-\omega_0t/2Q}\times\nonumber\\
&&\left[2Q\cos(\Omega\omega_0 t)+\Omega^{-1}\sin(\Omega\omega_0
t)\right]\bigg].
\end{eqnarray}
Recalling that $\alpha/\omega_0\gamma=Q$,
\begin{equation}\label{rt}
R_\theta(t)={kT\over
\alpha}e^{-\omega_0t/2Q}\Re\left[\cos(\Omega\omega_0 t)+(2\Omega
Q)^{-1}\sin(\Omega\omega_0 t)\right].
\end{equation}
The integral of the spectral density is $R_\theta(0)$ and should
be equal to the mean-square fluctuations,
\begin{equation}
R_\theta(0)={kT\over\alpha}
\end{equation}
which is true for Eq. (\ref{rt}) and is an important check for the
autocorrelation function.

The general operating procedure is to take many closely
time-spaced measurements of the torsion pendulum angle, with the
time spacing less than any correlation time in the system. The
variance of the sample mean for many closely spaced measurements
over a time interval $T_m$, assuming without loss of generality
that the mean is zero, is \cite{daven}
\begin{equation}
\sigma^2(\theta)={2\over T_m} \int_0^{T_m} \left(1-{t\over
T_m}\right)R_\theta(t) dt.
\end{equation}
In the limit of a very long measurement time, the second term in
the integral vanishes compared to the first, and in the limit of
large $Q$, we find
\begin{equation}\label{freevar}
\sigma^2(\theta)={2kT\over \alpha}{1\over Q\omega_0 T_m}.
\end{equation}
Noting that
\begin{equation} Q={\pi \tau_d \over\tau_0}={\omega_0
\tau_d\over 2}={I\omega_0\over\gamma}={\alpha\over\omega_0\gamma}
\end{equation}
where $\tau_0$ is the oscillation period and $\tau_d$ is the $1/e$
amplitude damping time, the variance can be recast in the form
\begin{equation}
\delta\theta_{\rm rms}=\sqrt{4kT\over I\tau_d
T_m}\left({\tau_0\over 2\pi}\right)^2
\end{equation}
which is a smaller by factor of $\sqrt{2/3}$ than the result given
in \cite{boynton}, and quoted in \cite{adelberger} as Eq. (10).

The force measurement noise is simply given by, using Eq.
(\ref{freevar}) and Eq. (6),
\begin{equation}
\delta F_{\rm rms}={\alpha\delta \theta_{\rm rms}\over
R}=\sqrt{2\gamma kT \over R^2 T_m}
\end{equation}
where $R$ is the effective radius where the force is applied.  We
thus see that $R$ should be as large as possible, $\gamma$ as
small as possible, and that the sensitivity is independent of
$\alpha$ and $I$.

\section{Torsion Pendulum with Electronic Feedback}

In the experiment described in \cite{lam}, feedback was used to
keep the torsion pendulum angle fixed in space.  The restoring
torque was generated by changing the voltage between a fixed plate
and the conducting pendulum body which was grounded.  The force
between capacitor plates with potential difference  $V$ is
proportional to $V^2$.  However, because the feedback signal was
added to a relatively large fixed voltage $V_0$, the system was
linear in that $V^2=(V_0+\delta V)^2\approx V_0^2 +2V_0\delta V$.

A simple proportional-plus-integral feedback scheme was used in
\cite{lam}.  For this system, Eq. (2) becomes
\begin{equation}
I\ddot \theta +\gamma\dot\theta +\alpha \theta
=\tau(t)-\beta\theta -\kappa\int \theta dt
\end{equation}
where $\beta$ and $\kappa$ are the proportional and integral
feedback gain, respectively.  The experiment \cite{lam} was
operated in the limit of very large damping which means that the
inertial ($I \ddot \theta$) term can be neglected.  The system was
in some ways equivalent to a phase-locked-loop (PLL) because in
this limit $\dot\theta$ is a constant value proportional to the
feedback signal \cite{gardner}, and the standard PLL analysis
techniques can be applied to this problem. The spectral amplitude
can be determined as before,
\begin{equation}
(i\gamma\omega + \alpha)\theta_\omega=\left(-\beta - {\kappa\over
i \omega}\right)\theta_\omega+\sqrt{4kT\gamma\Delta f}
\end{equation}
and the spectral density is
\begin{equation}
{\vert\theta_\omega\vert^2\over \Delta f}= {4kT\gamma\over
\kappa^2}{\omega^2\over (1-x^2)^2+Q^{-2}x^2}
\end{equation}
where
\begin{equation}
x={\omega\over \omega_0};\ \ \ \omega_0^2={\kappa\over \gamma};\ \
\ \ Q={\kappa\over \omega_0(\alpha+\beta)}.
\end{equation}
For this experiment, the output of the integrator provided the
torque measurement, and
\begin{equation}
{R^2\vert F_\omega\vert^2\over \Delta f}={\kappa^2\over
\omega^2}{\vert\theta_\omega\vert^2\over\Delta f}
\end{equation}
which is equal to Eq. (4) multiplied by $\alpha^2$, but with
different meanings for the parameters.  We can immediately write
down the force autocorrelation as measured at the integrator
output:
\begin{equation}
R_F(t)={\omega_0\gamma kT\over
R^2}e^{-\omega_0t/2Q}\left(Q\cos(\Omega\omega_0t)+(2\Omega)^{-1}\sin(\Omega\omega_0t)\right).
\end{equation}
The response time is optimum when $Q=0.5$ so that the system is
critically damped.  In this case it is easy to calculate the RMS
noise after many closely spaced measurements as before:
\begin{equation}\label{fn}
\delta F_{\rm rms}=\sqrt{2\gamma k T\over R^2T_m}
\end{equation}
which is the same result as before, Eq. (17).

\subsection{Electronic Noise}

So far in this analysis, electronic noise has been neglected. This
is reasonable because the differential capacitor signal used in
\cite{lam} to provide a measure of the torsion angle had an
intrinsic sensitivity of 100 V/rad. Given a typical integrated
operational amplifier noise of tens of nV/$\sqrt{\rm Hz}$ at the
AC brige frequency of 4.2 kHz (the amplifier noise voltage  is
comparable to the Johnson noise in the 1 M$\Omega$ bridge
resistors), the electronically limited angular resolution was of
order nRad/$\sqrt{\rm Hz}$, much below the intrinsic thermal
angular fluctuations of the torsion fiber, which, from Eq. (1)
were of order $\mu$Rad for $\alpha\approx 1$ dyne cm/Rad.

\subsection{Experimental examples}

Many recent measurements that have used torsion pendulums have
employed fine tungsten wires as the support and torsion element.
The tensile strength of Tungsten is among the highest of known
materials so a very fine diameter wire can be used to support a
substantial mass. Both the damping coefficient $\gamma_W$ and the
torque coefficient $\alpha$ depend on the radius to the fourth
power \cite{strong}:
\begin{equation}
\gamma_W={\eta\over 2}{\pi r^4\over \ell},\ \ \ \alpha={Z\over
2}{\pi r^4\over \ell}
\end{equation}
where $r$ is the wire radius, the internal viscosity
$\eta=9.37\times 10^{9}$ poise for tungsten (\cite{strong}, Table
Chap. V, Table III), and the tangential coefficient of elasticity
(or torsion modulus)  $Z\approx 1.8 \times 10^{12}$ dyne/cm$^2$,
and depends on the wire diameter (\cite{espe}, Fig. B 3.2-17). The
values of $\alpha$ for two experiments \cite{lam,adelberger} are
in reasonable agreement with this formula.

The experiments \cite{lam,adelberger} were operated with higher
damping than that intrinsic to tungsten wire, $\gamma_W$.  For
\cite{lam}, $\gamma_W=4.7\times 10^{-2}$ dyne cm/sec, compared to
the experiment $\gamma=10$ dyne cm/sec as set by the magnetic
damping. For \cite{adelberger}, $\gamma_W= 1.8 \times 10^{-4}$
dyne cm/sec, compared to the experiment $\gamma=0.2$ dyne cm/sec
due to background gas.  Both of these experiment relied on extra
damping to reduce the effects of perturbations that drove the
various pendulum modes of the system, and this extra damping
effectively increased the system bandwidth.

As we have shown in this note, the only parameter that affects the
force measurement sensitivity due to thermal fluctuations is the
mechanical damping coefficient $\gamma$. For the two experiments
discussed here, the wire length could have been substantially
shortened without increasing $\gamma$ significantly, e.g, for
\cite{lam} the intrinsic torsion wire damping becomes equal to the
magnetic damping when $\ell = 0.4 $ cm, and for \cite{adelberger},
equal to the residual gas damping when $\ell = 2.4$ cm. From a
thermal noise standpoint, there is very little gain in having the
wire length larger than these values.

The effects of tilt depend linearly on wire length, and tilt was a
major noise source in \cite{lam} where, for short-term noise,
thermal effects dominated. However, the $1/f$ noise corner was
around 5 Hz, and the effective system noise was about 100 times
larger than the expected thermal noise in the wire for
measurements periods of 10 seconds separated by 20 seconds. In
this case it was evident that the tilt was the largest noise
factor; distortion of the concrete floor by the weight of the
experimenter was a significant factor in the alignment of the
apparatus, and this floor was directly coupled to the building.
For \cite{adelberger}, the system noise was about 10 times the
thermal noise.  Effects of external noise due to tilt of the
apparatus could have been reduced, in a practical sense, by
reducing the torsion wire length by an order of magnitude in these
experiments without a significant loss of sensitivity.

\section{Conclusion: Application to Casimir Force Measurements}

We have shown that the single parameter that determines the
thermal force noise after a long-time average of a torsion
pendulum signal is the mechanical damping coefficient $\gamma$.
Several experiments that have been performed to date employed
extra system damping to shorten the time required to recover from
larger perturbations, and thereby effectively increase the system
bandwidth. This implies that many experiments could have employed
a shorter torsion wire, which would have substantially reduced the
noise due to tilting of the apparatus.

With electrical feedback, the system $Q$ can be arbitrarily tuned,
and in particular $Q=0.5$, corresponding to critical damping, can
be obtained.  Operating with this $Q$ provides the optimum
measurement bandwidth \cite{gardner}.  The thermal noise is still
determined by the intrinsic pendulum $\gamma$, see Eq. (\ref{fn}).
The implication is that the $Q$ can be made small, as is desirable
from bandwidth consideration (e.g., so that the system achieves
equilibrium in a shorter time after the force is changed).  A high
bandwidth allows the possibility of a measurement frequency higher
than the inevitable system $1/f$ noise corner frequency.

A short torsion wire (a few cm) suggests a new configuration for a
pendulum.  As shown in Fig. 1, the ``centers of force measurement"
at the test bodies can be located at the same height as the
support point of the torsion wire.  Then if the pendulum tilts,
the change in separation between the plates (in a Casimir
measurement) is second order in the tilt angle.  For flat plates,
there in only a change in separation when the tilt is
perpendicular to the axis formed by the line between the test
bodies. For curved plates, tilts around either axis lead to a
second-order in angle change in separation.  With such a
configuration, it should be possible to reduce the tilt noise by a
factor of 100, which would allow measurement at level determined
by the intrinsic thermal noise.  It is tempting to consider
contouring the plates so that there is no change in plate
separation with tilt angle, however, this possibility is likely
impractical.

The thermal correction to the Casimir force, between curved
perfectly conducting surfaces with net radius of curvature $\cal
R$, is \cite{mil}
\begin{equation}
F(d)\approx {2.4 {\cal R} kT\over 4 d^2}
\end{equation} where $d$
is the distance of separation between the curved surfaces, and
this equation is accurate for $d> 4\ \mu$m, where the usual
zero-point Casimir force is relatively small. The result in
\cite{bos} indicates a factor of two reduction in the force for
this or larger separations. In order to test this theory, a 10\%
measurement accuracy is needed at $d\approx 4 \ \mu$m, or $\delta
F=0.1\ \mu$dyne for ${\cal R}=10$ cm.  This level of sensitivity
is 10 times the intrinsic thermal noise of \cite{lam}, so with
control of tilt noise, a relevant test of the theory \cite{bos}
should be possible.  The idea put forward in \cite{lam2} is to
employ materials with different electron mean free paths compared
to the electromagnetic skin depth, for mode frequencies that
contribute significantly to the Casimir force. The analysis
\cite{bos} is only applicable for materials where the skin depth
is greater than the mean free path. In the opposite case, the
surface impedance can be used to calculate the force
\cite{mostep,lam2} and the perfectly conducting thermal correction
should be obtained. Germanium infrared optical lenses are
available with $\cal R$ of order 10 cm \cite{ealing}, and these
can be used as the plates in a Casimir force measurement using a
torsion pendulum as in \cite{lam}.

\begin{figure}
\begin{center}
\includegraphics[
width=3.5in ] {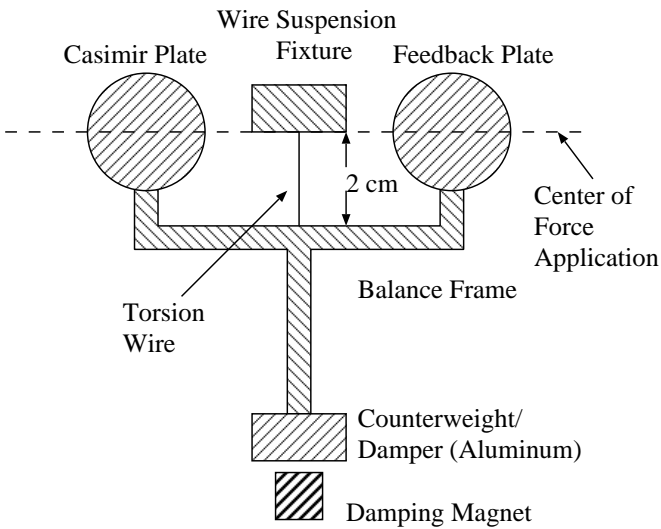} \caption{In addition to shortening
the wire length in a torsion pendulum, placing the centers of the
electrostatic feedback and Casimir plates at the point of
suspension of the torsion wire further reduces the effects of
apparatus tilt.}
\end{center}
\end{figure}

\end{document}